\documentclass[aps,pra,twocolumn,nofootinbib]{revtex4-1}
\newcommand {\be}{\begin{equation}}
\newcommand {\ee}{\end{equation}}
\newcommand {\bey}{\begin{eqnarray}}
\newcommand {\eey}{\end{eqnarray}}

\usepackage{epsfig}
\usepackage{amsfonts}
\usepackage{amsmath}
\usepackage{mathbbol}
\usepackage{enumerate}
\usepackage{amsthm}
\usepackage{bbm}
\usepackage{pgfplots}
\usepackage{tikz}
\usepackage[utf8]{inputenc}

\usepackage[colorlinks=true,citecolor=black,urlcolor=black,linkcolor=black]{hyperref}

\newcommand{\tr}[1]{\operatorname{tr}\!\left[#1\right]}

\theoremstyle{definition}

\let\originaleqref\eqref
\renewcommand{\eqref}{Eq.~\originaleqref}

\usepackage[pangram]{blindtext}
\usepackage{balance}
\usepackage{lastpage}

\begin{document}

\title{Preliminary investigation of a new state estimation technique using device-independent maximum likelihood estimation}
\author{Sacha Schwarz$^{1,2,3}$, Bruno Eckmann$^{1,4}$, Denis Rosset$^{5,6}$, and Andr\'e Stefanov$^1$}
\affiliation{${}^1$Institute of Applied Physics, University of Bern, 3012 Bern, Switzerland \\
${}^2$Institute for Quantum Computing, University of Waterloo, Waterloo, Ontario, Canada, N2L 3G1\\
${}^3$Department of Physics \& Astronomy, University of Waterloo, Waterloo, Ontario, Canada, N2L 3G1\\
${}^4$Institute for Theoretical Physics, ETH Zurich, 8093 Zurich, Switzerland \\
${}^5$Department of Physics, National Cheng Kung University, Taiwan 701, Taiwan \\
${}^6$Perimeter Institute for Theoretical Physics, Waterloo, Ontario, Canada, N2L 2Y5}
%\date{\today}

\begin{abstract}
  In quantum state tomography, the estimated frequencies do not correspond directly to a physical quantum state, due to statistical fluctuations.
  Thus, one resorts to point estimators that return the state that matches observations the best, and a variety of estimators have been proposed -- linear inversion, least squares, maximum likelihood (ML) -- each making different trade-offs.
  In this short note, we investigate an alternative approach  inspired by device-independent quantum information protocols.
  We embed a tomographic complete set of measurement operators within the framework of a Bell scenario, and first regularize the estimated frequencies using a device-independent maximum likelihood algorithm.
  We then run the standard maximum likelihood algorithm to estimate the underlying quantum state.
  In this preliminary investigation, we do not observe significant differences in the reconstructed state when using the new algorithm.
\end{abstract}

\maketitle

In many different experimental studies of quantum mechanical systems, a detailed knowledge of the prepared quantum state is advantageous for further processing tasks \cite{White1999,Mair2001,Marcikic2001,Agnew2011,Schwarz2014}.
Therefore, Quantum State Tomography (QST) is usually employed in order to fully reconstruct the corresponding density matrix \cite{Agnew2011,James2001,Thew2002}.
A finite tomographic complete set of projective measurement operators is applied onto the experimentally prepared quantum system and the density matrix estimate is derived through linear inversion (LIN) of the measurement results.
But, due to the limitation to finite statistics, the estimated state is typically not physical.
This is why reconstruction schemes like Maximum-Likelihood (ML) estimations were introduced \cite{James2001}.
There, a density matrix is reconstructed by maximizing some target function within the set of physical density operators.
It is however known that in such parameter estimations for finite statistics, systematic deviations occur which are finally leading to biased estimators.
In general, every density matrix reconstruction scheme, which is constrained to yield a physical density matrix, leads always to a biased estimator \cite{Schwemmer2015}.
Hence, QST has to deal with the trade-off between non-physical but unbiased reconstructed density matrices by linearly inverting the measurement results, and physical but biased density matrices when we apply reconstruction schemes like ML.
It is therefore relevant to investigate methods which minimize both the bias and the non-physicality of the reconstructed states.

In device-independent (DI) quantum information, a similar problem exists.
Due to statistical fluctuations, DI analysis tools are usually not directly applicable onto raw data sets, because they generally do not satisfy the conditions of no-signaling \cite{Schwarz2016}.
In DI assessments, no assumptions are made on the experimental apparatus, nor on the explicit quantum state dimension, such that any conclusion is deduced uniquely from the measurement statistics.
However, to bridge the gap between the theoretical DI tools and experimentally determined data, a similar idea as in ML estimation has been published recently in~\cite{Lin2017}, proposing a regularization algorithm of raw correlation data sets within the framework of a Bell experiment.

In this paper, we investigate different variants of regularization schemes to explore alternative approaches to ML estimations.
We construct a new approach, where the quantum state is not directly reconstructed from raw data sets but from some regularized input-output probability estimates.
In the context of a Bell scenario, a set of projective measurement operators are thereby constructed such that a tomographic complete set of measurement operators is incorporated.
We focus on the density matrix reconstruction of entangled bipartite quantum states as an exemplary system.
Based on the fidelity, a commonly used quantum state characterization measure, we compare those different approaches.
Comparing the new method introduced with standard maximum likelihood estimation, we do not find a significant improvement on this figure of merit.
In addition, the reconstructed states are similar between the two methods; as measured by the trace distance, the difference is an order of magnitude smaller than the 
We do not observe a significant improvement on this figure of merit 
Additionally, we measure the difference between those two methods using the trace distance of reconstructed states: typically, this difference is around an order of magnitude smaller than the distance with respect to the actual underlying state.

\section{Quantum tomography}

We consider a bipartite Bell scenario with three inputs and two outputs incorporating qubits. In such a scenario, we have two parties, usually denoted as Alice ($\text{A}$) and Bob ($\text{B}$), who receive one particle each. Both particles originate from the same quantum source emitting particle pairs and are described by a two-qubit density matrix $\rho$. Alice and Bob are then able to perform local measurements, i.e., Alice performs some measurement $M_{a|x}^\text{A}$ and Bob some measurements $M_{b|y}^\text{B}$, where we denote the input settings as $x,y \in \{1,2,3\}$ and the output labels as $a,b \in \{0,1\}$. We express the local measurements as
\begin{equation}
  M_{a|x}^\text{A} = \frac{\mathbbm{1} + (-1)^a \sigma_x }{2}\;, \quad M_{b|y}^\text{B} = \frac{\mathbbm{1} + (-1)^b \sigma_y }{2}\;,
  \label{eq:localMeasurement}
\end{equation}
where we make use of the Pauli matrices $\sigma_n$ with $n \in \{1,2,3\}$. Further, we assume that the inputs are chosen uniformly and at random, according to a distribution $P(xy) = 1/9$.
Incorporating the selection of inputs inside the POVM, we get
\begin{equation}
  \label{eq:ourmeasurements}
  M_{abxy} = P(xy) \left( M_{a|x}^\text{A} \otimes M_{b|y}^\text{B} \right)\;.
\end{equation}

The corresponding input-output correlations in such a scenario are characterized by joint probabilities $P(a b x y)$:
\begin{equation}
  \label{eq:measresult}
  P(a b x y) = \tr{ M_{abxy} ~ \rho}\;,
\end{equation}
 which are commonly summarized by the probability vector $\vec{P}=\{ P(a b x y)\}_{a,b,x,y}$.

In experiments, joint conditional probabilities of the form $P(a b x y)$ are estimated by relative frequencies $f(ab xy)$ deduced from finite measurement statistics. These quantities are computed via the number of coincidences $N(abxy)$, measured for a particular combination of inputs $x, y$ and outputs $a, b$, i.e.,
\begin{equation}
P(a b x y) \approx f(abxy) = \frac{N(abxy)}{N}, \quad N = \sum_{abxy} N(abxy)\;,
\label{eq:relativeFrequencies}
\end{equation}
where $N$ is the total number of events.
For our purposes, we assume a Poissonian distribution of coincidence detection rates at a constant efficiency $\eta$ with an average $\langle N \rangle = \eta T$, where $T$ is the total measurement time.

\subsection{Linear inversion for state estimation}

A two-qubit density matrix $\rho$ can be expressed as a linear combination of Pauli matrices $\sigma_i$ and the $2$-dimensional identity matrix which we write $\sigma_0 = \mathbb{1}_2$~\cite{James2001,Schwemmer2015}.
Thus, we write 
\begin{equation}
\label{eq:singleQubitDensityMatrix}
\rho = \frac{1}{4}\sum_{jk=0}^{3} T_{jk} ~ \Gamma_{jk}, \qquad \Gamma_{jk} = \big(\sigma_j \otimes \sigma_k \big) \;,
\end{equation}
where $T_{jk} = \tr{\rho \, \Gamma_{jk}}$, as $\tr{\Gamma_{jk} ~ \Gamma_{lm}} = 4 ~ \delta_{jl} ~ \delta_{km}$.
To simplify the notation, we identify the indices
\[
  (j,k) = (0,0), (0,1), \ldots, (0,3), \ldots, (3,3)\;,
\]
with $i = 1,\ldots,16$.

Now, consider an informationally complete measurement described by the POVM elements $\{ M_\mu \}$ for $\mu = 1, \ldots, n_M$, not necessarily (yet) the measurement described in~\eqref{eq:ourmeasurements}.
The probability of getting the measurement outcome $\mu$ is given by
\begin{equation}
  \label{eq:ProbOfMeasurement}
  P_\mu = \tr{\rho ~ M_\mu} = \sum_{i=1}^{16} \tr{M_\mu ~ \Gamma_i} ~ T_i\;.
\end{equation}

By defining the matrix $\hat{B}$ to be composed with matrix elements
\begin{equation}
\label{eq:Boperator}
\hat{B}_{\mu,i} = \frac{1}{4} \tr{M_\mu \, \Gamma_i}\;,
\end{equation}
the expression in \eqref{eq:ProbOfMeasurement} simplifies to
\begin{equation}
  \label{eq:linearsystem}
  \vec{P} = \hat{B} ~ \vec{T}\;.
\end{equation}

Now, when replacing the ideal probability distribution $\vec{P}$ by the estimated frequencies $\vec{f}$ 
a linear system that can in principle be solved to recover $\vec{T} = \hat{B}^{-1} ~ \vec{f}$ from the frequencies $\vec{f}$, and thus obtain an estimation of $\rho$.
This method is called {\em linear inversion}.

In the case where the POVM has exactly $n_M = 16$ outcomes, the linear system in~\eqref{eq:linearsystem} always has a single solution; for the measurements we defined in~\eqref{eq:ourmeasurements}, the system is overcomplete, and thus likely to be inconsistent in the presence of statistical noise.
However, if we replace the (then) undefined inverse $\hat{B}^{-1}$ by the Moore-Pensore~\cite{Golub2012} pseudoinverse $\hat{B}^+$ so that $\vec{T} = \hat{B}^+ ~ \vec{f}$; in that case the solution corresponds to the argument $\vec{T}$ minimizing the following objective:
\begin{align*}
  \min_{\vec{T}} & \left \Vert \vec{P} - \vec{f} \right \Vert_2 \\
  \text{so that } \vec{P} & = \hat{B} ~ \vec{T}\;.
\end{align*}

Still, there is no guarantee that the resulting density matrix $\rho$ is physical: it could well have negative eigenvalues.

\subsection{Maximum likelihood estimation}
To solve these problems, the maximum likelihood estimation (MLE) method is commonly employed~\cite{James2001}. 
Instead of minimizing the 2-norm, it minimizes instead the Kullback-Leibler divergence~\cite{Kullback1951} (for a statistical motivation, see~\cite{VanDam2005}), defined as
\[
  D(\vec{f} \Vert \vec{P}) = \sum_i f(i) \log \frac{f(i)}{P(i)}\;.
\]
In addition, it requires that $\vec{P}$ corresponds to a normalized, semidefinite positive solution of~\eqref{eq:measresult} -- that is $\tr{\rho} = 1$ and $\rho \ge 0$.
This optimization can be cast in a conic linear optimization problem, expressed with semidefinite and exponential cones.
However, few solvers~\cite{SCS} support that mixture of cones; instead, we use the diluted iterative algorithm presented in~\cite{Rehacek2007}.
The iteration is defined using the state-dependent operator
\[
  \hat{R}(\hat{\rho}) = \frac{1}{N} \sum_i \frac{f(i)}{\hat{P}(i)} M_i\;,
\]
where $\hat{\rho}$ is the value of the reconstructed state (which depends on the current iteration), and the current reconstructed statistics are $\hat{P}(i) = \tr{\hat{\rho} ~ M_i}$.
We then compute the state for the next iteration as
\[
  \hat{\rho}' = \mathcal{N}\left[ (\mathbbm{1} + \varepsilon \hat{R}) \hat{\rho} (\mathbbm{1} + \varepsilon \hat{R}) \right], \quad \mathcal{N}(\tau) = \frac{\tau}{\tr{\tau}}\;,
\]
where $\varepsilon$ is the dilution parameter.
We start with $\varepsilon = 10^6$, and set $\varepsilon' = \varepsilon / 10$ when the iteration does not improve the objective.

\subsection{Observing a subset of events}
In order to form a tomographically complete set of measurements as chosen in~\cite{James2001}, not all events have to be observed in the chosen Bell scenario with three binary-outcome measurements.
By only keeping the measurement outcomes $a,b = 0,1$ for the measurements settings $x,y = 1$  and $a,b=0$ in the case of $x,y = 2,3$, we are not considering the full set of counts $N(abxy)$ than rather a subset of $(a,b,x,y) \in \mathcal{I}$.
In that case, the frequencies cannot be directly estimated, as the total number of events is unknown.
We now assume that there is a subset $\mathcal{J} \subset \mathcal{I}$, such that the sum of the corresponding POVM elements is proportional to the identity:
\[
  M_{\mathcal{J}} = \sum_{(a,b,x,y)\in\mathcal{J}} M_{abxy} = \alpha ~ \mathbbm{1}\;.
\]
Then, we estimate the total number of events
\[
  \hat{N} = \sum_{(a,b,x,y)\in\mathcal{J}} N_{abxy} / \alpha\;,
\]
and estimate the frequencies accordingly.

The methods above can still be used by considering a restricted POVM with elements
\[
  \{ M_{abxy} : (a,b,x,y) \in \mathcal{J} \} \cup \{ M_{\overline{\mathcal{J}}} \} \;, \quad M_{\overline{\mathcal{J}}} = \mathbbm{1} - M_{\mathcal{J}}\;,
\]
and estimating the frequency of unobserved events by
\[
  f_{\overline{\mathcal{J}}} = 1 - \sum_{(a,b,x,y)\in\mathcal{J}} f(abxy)\;,
\]
provided this restricted POVM is still informationally complete.

\section{Device-independent methods applied to tomography}

We now investigate the use of device-independent (DI) preprocessing methods in the context of quantum state tomography.
In that context, we consider the correlations obtained when devices are given external inputs $x,y$ and produce outputs $a,b$; the resulting behavior is described using the conditional distribution
\[
  P(ab|xy) = \tr{(M^\text{A}_{a|x}\otimes M^\text{B}_{b|y}) ~ \rho}\;.
\]
No assumptions are made about the POVM elements $\{ M^\text{A}_{a|x} \}$ and $\{ M^\text{B}_{b|y} \}$.
Compared to device-dependent tomographic reconstruction, remark that $P(ab|xy)$ is now a {\em conditional} probability distribution, which we collect in a vector $\vec{P}_\text{cond}$.

Consequently, we say that behavior $\vec{P}_\text{cond}$ belongs to the set of all possible quantum correlations $\mathcal{Q}$, i.e., $\vec{P}_\text{cond} \in \mathcal{Q}$.
Every behavior $\vec{P}_\text{cond} \in \mathcal{Q}$ satisfies the nonsignaling conditions
\begin{equation}
\begin{aligned}
P(a \vert xy) &\equiv \sum_b P(a b \vert x y) = P(a \vert x y')\;, \quad \forall a,x,y,y'\;, \\
P(b \vert xy) &\equiv \sum_a P(a b \vert x y) = P(a \vert x' y)\;, \quad \forall a,x',y,y\;,
\end{aligned}
\label{eq:nonsignalingConditions}
\end{equation}
such that we have the inclusion $\mathcal{Q} \subset \mathcal{NS}$.
This follows from the fact that the behaviors $\vec{P}_\text{cond}$ satisfying \eqref{eq:nonsignalingConditions}, form themselves a set, namely the set of nonsignaling correlations $\mathcal{NS}$, which incorporates the set $\mathcal{Q}$ entirely \cite{Brunner2014}.
Quantum behaviors are fully described by the hierarchy of semidefinite relaxations proposed in~\cite{Navascues2008a}.
At finite levels in the hierarchy, these relaxations are outer approximations of the quantum set.
Nevertheless, the resulting approximations are useful in a variety of contexts, for example entanglement estimation~\cite{Lin2017} and randomness extraction~\cite{Bourdoncle2018a}.

In particular, given the counts of observed events $N(abxy)$, one can ask the most likely underlying approximate quantum behavior.
In~\cite{Lin2017}, a method inspired by MLE state reconstructed was proposed: then, one estimates $P(ab|xy)$ by the conditional frequencies
\[
  f(ab|xy) = \frac{N(abxy)}{N(xy)}
\]
and then minimizes the KL divergence between $\vec{f}_\text{cond}$ and a behavior $\vec{P}_\text{DI} \in \tilde{\mathcal{Q}}$ contained in a relaxation $\tilde{\mathcal{Q}} \supset \mathcal{Q}$.
The resulting behavior $\vec{P}_\text{DI}$ then corresponds to the regularization of the frequencies $\vec{f}_\text{cond}$ with respect to the quantum relaxation $\tilde{\mathcal{Q}}$.
In the present study, we use the level with the set of monomials $\left \{\mathbbm{1}, M_{a|x}^\text{A}, M_{b|y}^\text{B}, M_{a|x}^\text{A} M_{b|y}^\text{B} \right \}$.

Inspired by the resulting obtained by device-independent regularization, the state reconstruction method we study is the following.
Given the event counts $N(abxy)$, we estimate the most likely approximate quantum behavior $\vec{P}_\text{DI}$, without making assumptions about the measurements performed, by minimizing the KL divergence $D(\vec{f}_\text{cond}\Vert \vec{P}_\text{DI})$.
Note that $\vec{P}_\text{DI}$ is a conditional distribution.
We obtain the unconditional distribution by multiplying back with the estimated input frequencies: 
\[
  P_\text{uncond}(abxy) = P_\text{DI}(ab|xy) f(xy)
\]
and then use the regularized $\vec{P}_\text{uncond}$ to minimize the KL divergence~\cite{VanDam2005,Acin2005} $D(\vec{P}_\text{uncond}\Vert \vec{P}_\text{DD})$ between this regularized behavior and the underlying distribution $\vec{P}_\text{DD}$ obtained with the known, trusted measurements $M_{abxy}$ of~\eqref{eq:ourmeasurements}.
We finally reconstruct the state using the behavior $\vec{P}_\text{DD}$, which is by construction consistent with the tomography equation~\eqref{eq:ProbOfMeasurement} -- whereas $\vec{P}_\text{uncond}$ is not necessarily consistent with it.

\section{Benchmark of state reconstruction methods}

\begin{figure}
  \includegraphics{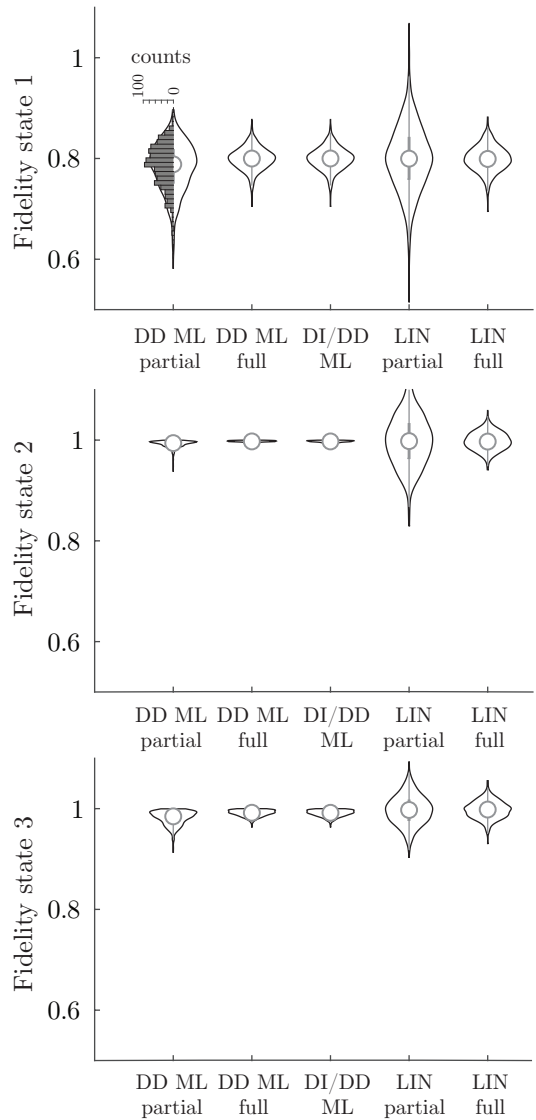}
  \caption{
    Estimated fidelities for $R=1000$ runs, each including $N=1000$ samples; for lisibility, the data is presented using violin plots, and for the first dataset we superimposed the underlying histogram.
    There are three graphs for the states $\tau_1$, $\tau_2$, $\tau_3$; the $x$ axis presents the five methods investigated in the present work (see main text).
    In this particular example, we see that the hybrid DI/DD method does not provide an advantage over standard state reconstruction techniques; and that linear inversion can estimate nonphysical density matrices, as evidenced by fidelities greater than one.
  }
  \label{fig:fidelityComparison}
\end{figure}

We investigated the five methods outlined above, namely:
\begin{itemize}
\item {\em DD ML partial} estimates the state using diluted maximum likelihood iterations using a subset of events;
\item {\em DD ML full} uses the full dataset with diluted maximum likelihood estimation;
\item {\em DI/DD ML} uses the hybrid method that combines device-independent and device-dependent maximum likelihood estimation;
\item {\em LIN partial} performs linear inversion on the estimated frequencies for a subset of events;
\item {\em LIN full} perform linear inversion on the full tomographic data, minimizing the $2$-norm of the residual.  
\end{itemize}`

We used three states as candidates.
The test states $\tau_1$ and $\tau_2$ are obtained from the singlet state 
\begin{equation*}
\vert \psi \rangle = \frac{1}{\sqrt{2}}\left( \vert 0 \rangle_A \vert 0 \rangle_B + \vert 1 \rangle_A \vert 1 \rangle_B \right)
\end{equation*}
to which we add the amounts $\lambda_1 = 0.52$ and $\lambda_2 = 0.995$ of white noise, so that:
\begin{equation*}
\tau_i = \lambda_i \vert \psi \rangle \langle \psi \vert + \frac{1-\lambda_i}{4}\mathbb{I}_4\;.
\end{equation*}
For these two states, the resulting fidelity is given by
\[
  \mathcal{F}(\hat{\rho}, \vert\psi\rangle) = \langle\psi\vert \hat{\rho} \vert\psi\rangle\;,
\]
where $\hat{\rho}$ is the estimated density matrix.
The third test state $\tau_3$ is obtained by adding white noise to the partially entangled state
\begin{equation*}
\vert \phi \rangle =\alpha \vert 0 \rangle_A \vert 0 \rangle_B + \beta \vert 1 \rangle_A \vert 1 \rangle_B
\end{equation*}
where $\alpha = 0.961$ and $\beta = 0.276$, so that
\begin{equation*}
\tau_3 = \lambda_3 \vert \phi \rangle \langle \phi \vert + \frac{1-\lambda_3}{4}\mathbb{I}_4
\end{equation*}
with $\lambda_3 = 0.995$, and the fidelity is given by
\[
  \mathcal{F}(\hat{\rho}, \vert\phi\rangle) = \langle\phi\vert \hat{\rho} \vert\phi\rangle\;.
\]

We present the results of these preliminary investigations in Figure~\ref{fig:fidelityComparison}.
We include simulations using subsets of observed events as a point of comparison; however the hybrid {\em DI/DD ML} needs the full dataset to be employed
\footnote{
  A previous version of this manuscript showed an improvement for the {\em DI/DD ML full} method compared to the {\em DD ML partial} method, with a longer simulated running time for the partial dataset.
  This effect was an artifact of the methodology, and disappears in the current investigation.
}.
On the full dataset, we see that the {\em DI/DD ML} method we introduce gives similar results when compared to the standard {\em DD ML} method.

\begin{figure}
  \includegraphics{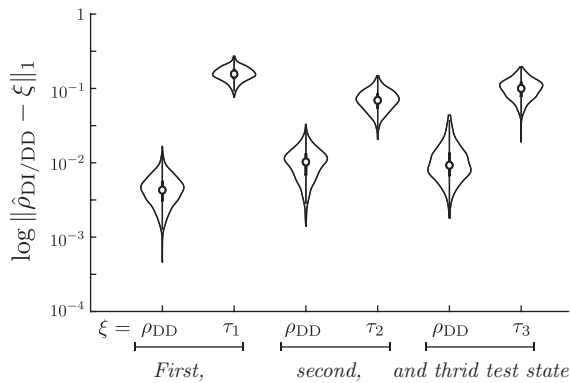}
  \caption{
    Estimated trace distances for $R=1000$ runs, using the same dataset as in Figure~\ref{fig:fidelityComparison}.
    We compare the trace distance between the states reconstructed with the two methods, and the trace distance to the underlying state.
  }
  \label{fig:traceDistance}
\end{figure}
Due to the similarity in the fidelity figure, we verified how similar the reconstructed states were for the {\em DI/DD ML} method ($\hat{\rho}_\text{DI/DD}$) and the standard {\em DD ML} method ($\hat{\rho}_\text{DD}$).
For that purpose, we compute the trace distance between the reconstructed DI/DD state and the state $\xi$
\[
  \mathcal{D}(\hat{\rho}_\text{DI/DD}, \xi) = \Vert \hat{\rho}_\text{DI/DD} - \xi \Vert_1
\]
where $\xi$ is either the state $\hat{\rho}_\text{DD}$ reconstructed with the standard ML technique, or the target state $\tau_i$.
The results are displayed in Figure~\ref{fig:traceDistance}.
We see that while the two methods do not reconstruct the same state, the differences are an order of magnitude smaller than the statistical error itself.

\section{Conclusion}
We studied the behavior of different state reconstruction methods.
In particular, we explored whether device-independent estimation techniques could improve standard point estimators.
Our new approach is based on successive maximum likelihood estimation, introducing an additional step that regularizes the correlations with respect to device-independent constraints.
In comparison with standard maximum likelihood, our preliminary investigation does not show a significant improvement on the fidelity of the reconstructed state; moreover, the states estimated by the two methods are pretty similar.
We leave as an open question whether this new technique can improve upon standard MLE with respect to different figures of merit or in different scenarios.

\begin{acknowledgments}
  We thank Yeong-Cherng Liang for useful discussions and comments.
  This research was supported in part at the Institute of Applied Physics by the Swiss National Science Foundation (Grant No. PP00P2\_159259), and at Perimeter Institute and the Institute for Quantum Computation by the Government of Canada through Industry Canada and by the Province of Ontario through the Ministry of Research and Innovation. This publication was made possible through the support of a grant from the John Templeton Foundation.
\end{acknowledgments}

\end{document}